\documentclass[prl,floatfix,showpacs,amsmath,twocolumn]{revtex4}
\usepackage{amssymb}
\usepackage{graphics}
\usepackage{epsfig}
\usepackage{graphicx}
\usepackage{psfrag}

\begin{document}

\newcommand{\be}{\begin{equation}}
\newcommand{\ee}{\end{equation}}
\newcommand{\bea}{\begin{eqnarray}}
\newcommand{\eea}{\end{eqnarray}}
\newcommand{\bma}{\begin{subequations}}
\newcommand{\ema}{\end{subequations}}
\def\lR{l^2_{\mathbb{R}}}
\def\RR{\mathbb{R}}
\def\E{\mathbf e}
\def\D{\boldsymbol \delta}
\def\S{{\cal S}}
\def\T{{\cal T}}
\def\dd{\delta}
\def\one{{\bf 1}}
\def\ss{\boldsymbol \sigma}

\newtheorem{theorem}{Theorem}
\newtheorem{lemma}{Lemma}
\newtheorem{definition}{Definition}

\title{Diverging Entanglement Length in Gapped Quantum Spin Systems}

\author{F. Verstraete$^{1}$, M.A.
Mart\'{\i}n-Delgado$^{2}$, and J.I. Cirac$^{1}$} \affiliation{(1)
Max-Planck-Institut f\"ur Quantenoptik,
Hans-Kopfermann-Str. 1, D-85748 Garching, Germany.\\
(2) Departamento de F\'isica Te\'orica I, Universidad Complutense
de Madrid, E-28040, Spain}

\date{\today}

\begin{abstract}
We prove the existence of gapped quantum Hamiltonians whose ground
states exhibit an infinite entanglement length, as opposed to
their finite correlation length. Using the concept of entanglement
swapping, the localizable entanglement is calculated exactly for
valence bond and finitely correlated states, and the existence of
the so--called string-order parameter is discussed. We also report
on evidence that the ground state of an antiferromagnetic chain
can be used as a perfect quantum channel if local measurements on
the individual spins can be implemented.
\end{abstract}

\maketitle

The fields of Condensed Matter and Quantum Information Theory
share a common interest in the study of quantum states of
many--body systems. Much of the current effort in Quantum
Information Theory is devoted to the description and
quantification of the entanglement contained in quantum states in
general: this intriguing property of Quantum Mechanics is the
basic resource of most of the applications in this field,
including quantum communication and computation. Condensed matter
theory, on the other hand, is deeply interested in the strongly
correlated states appearing in certain materials at very low
temperatures, since they describe a variety of fascinating
phenomena, like the ones occurring in quantum phase transitions
and superconductivity.

Despite the fact that the number of parameters to describe a
quantum state scales exponentially in the number of particles, it
is sometimes possible to capture the most relevant physical
properties by describing these systems in terms of very few
parameters. In the case of spin chains, for example, two--particle
correlations play a fundamental role. They allow us to understand
several complex physical phenomena, like phase transitions and the
appearance of a length scale in the system. Much insight has also
been obtained by studying exactly solvable models such as the
AKLT-model \cite{AKLT}, which illustrates the appearance of the
Haldane gap \cite{Haldane} in spin--1 antiferromagnets and the
associated finite correlation length.

From the perspective of entanglement theory, the presence of
two-particle correlations between two distant particles in a
many-particle pure state guarantees the possibility of
establishing EPR-type entanglement between them by doing local
measurements on the other particles \cite{VPC03}. On the other
hand, highly entangled multiparticle pure states typically have
reduced two-particle density operators close to the maximally
mixed state and therefore only exhibit very small correlations.
Correlation functions can therefore only reveal partial
information about the long-range quantum correlations that ought
to be present in a state. The so--called Localizable Entanglement
(LE) \cite{VPC03} between two particles is defined as the maximal
possible bipartite entanglement that can be \emph{localized}
between them, on average, by optimizing over all possible local
measurements on the other particles. The LE has a very clear
operational meaning as it quantifies the amount of entanglement
that can be localized at e.g. the end points of a spin chain and
that could be used, e.g., as a perfect quantum channel. Just as
correlation functions induce a correlation length $\xi_C$ in a
lattice, the LE induces a new length scale which we call the
entanglement length $\xi_E$. It has been proved in \cite{VPC03}
that $\xi_E\geq\xi_C$; therefore a diverging correlation length at
e.g. a quantum phase transition implies a diverging entanglement
length. In the case of ground states of typical spin--$1/2$
systems such as the Ising chain and the Heisenberg antiferromagnet
in a magnetic field, we observed that the bound is tight
$\xi_E=\xi_C$ \cite{Popp} and hence the opposite also holds true
in this case. This triggered the quest for a phase transition that
is detected by a diverging $\xi_E$ but for which the $\xi_C$
remains finite. The natural candidates are the spin--1
Hamiltonians that have a Haldane gap \cite{Haldane} and hence
finite correlation length.

In this paper, the following results are established: 1/ We show
that for a family of interesting spin--1 Hamiltonians, including
the celebrated AKLT-model, the entanglement length diverges as
physical parameters are changed, whereas the correlation length
remains finite. 2/ We calculate the LE exactly for a whole class
of finitely correlated states \cite{FNW}, which are
generalizations of the AKLT-ground state. This is a highly
nontrivial result as the definition of the LE involves a complex
optimization over all possible measurement strategies. 3/ The
so--called string-order parameter \cite{string}, reflecting a
mysterious topological hidden long-range order in spin--1
antiferromagnets, is given a natural interpretation in terms of
the LE. It is shown that, in the family of deformed AKLT-models,
there exist ground states with infinite $\xi_E$ but vanishing
string order parameter. 4/ We report on numerical results
indicating that the LE between the two end points of a Heisenberg
spin--1 antiferromagnetic chain is the maximal possible one. These
results indicate that ideas and techniques developed in the last
few years in the field of Quantum Information Theory prove useful
to analyze and understand certain aspects of the field of
Condensed Matter (see also \cite{preskill}).

\begin{figure}[t]
\psfrag{c}[Bc][Bc][1][0]{a)} \psfrag{d}[Bc][Bc][1][0]{b)}
\psfrag{a}[Bc][Bc][0.75][0]{$\bar{0}$}
\psfrag{b}[Bc][Bc][0.75][0]{$N+1$}
\psfrag{I}[Bc][Bc][0.75][0]{$|I\rangle$}
\psfrag{p}[Bc][Bc][0.75][0]{$|\phi\rangle$}
\psfrag{1}[Bc][Bc][0.75][0]{$1$}
\psfrag{2}[Bc][Bc][0.75][0]{$\bar{1}$}
\psfrag{3}[Bc][Bc][0.75][0]{$2$}
\psfrag{4}[Bc][Bc][0.75][0]{$\bar{2}$}
\psfrag{5}[Bc][Bc][0.75][0]{$3$}
\psfrag{6}[Bc][Bc][0.75][0]{$\bar{3}$}
\psfrag{x}[Bc][Bc][0.75][0]{$\ldots$}
\includegraphics[scale=0.40]{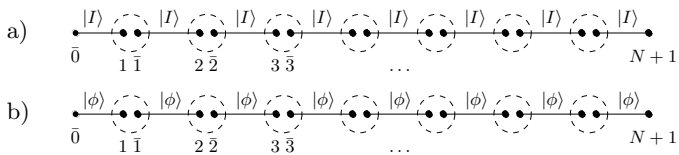}
\caption{(a) The Valence Bond ground state of the AKLT-model
(\ref{AKLT}); each edge represents a singlet, and the dashed
circle corresponds to a projection onto the symmetric subspace.
(b) A deformed AKLT model where the singlets are replaced by
non-maximally entangled states $|\phi\rangle$.} \label{fig0}
\end{figure}

Let us start by recalling the AKLT-model \cite{AKLT}, which plays
a central role in the understanding of gapped spin systems. To
make calculations simpler, we will consider an open chain of $N$
spin--1 particles at positions $1,2,\ldots,N$, and with
spin--$1/2$ particles at the ends (locations $0$ and $N+1$). The
AKLT-Hamiltonian is \be\label{AKLT} H^{\rm AKLT}:=\sum_{k=0}^N
X^{\rm AKLT}_{k,k+1}:=\sum_{k=0}^N\left(
\vec{S}_k\vec{S}_{k+1}+\frac{1}{3}(\vec{S}_k\vec{S}_{k+1})^2+\frac{2}{3}\right)
\ee with $\vec{S}$ the three spin operators. Each term $X^{\rm
AKLT}_{k,k+1}$ is a projector on a 5-D subspace of the 9-D space
of two spin--1 particles. The ground state can be obtained by
representing each spin--1 ($1\ldots N$) by $2$ spin--$1/2$'s
($\bar{1},1,\bar{2},2,\ldots$) and project them on the 3-D
symmetric subspace (see Fig.~\ref{fig0}). It can indeed be checked
that $X^{\rm AKLT}$ (\ref{AKLT}) is orthogonal to all operators of
the form \be\label{proj}A_{\bar{1}1}\otimes
A_{\bar{2}2}[X_{\bar{1}}\otimes |I\rangle_{1\bar{2}}\langle
I|\otimes X_{2}]A_{\bar{1}1}^\dagger\otimes
A_{\bar{2}2}^\dagger\ee with $X_{\bar{1}},X_2$ arbitrary $2\times
2$ operators, $|I\rangle$ the singlet of two qubits
$|I\rangle=|01\rangle-|10\rangle$, and $A$ the $3\times 4$
operator that projects a system of two qubits onto its symmetric
subspace. The ground state of $H^{\rm AKLT}$ is unique \cite{AKLT}
and can be written as
 \be \label{VBS0}
 |V\rangle = \left(\otimes_{k=1}^N
 A_{\bar{k}k}\right)|I\rangle_{\bar{0}1}|I\rangle_{\bar{1}2}\cdots
 |I\rangle_{\bar{N}N+1}.\ee
Indeed, all reduced density operators of two nearest neighbor
spins of $|V\rangle$ are of the form (\ref{proj}).

Due to the special structure of this state, the following basic
properties of singlets can be used to get some insight about the
quantum correlations present: i/ Given a complete spin-1 basis
$|\beta\rangle$ and a $3\times 4$ operator $A$, then there always
exist $2\times 2$ operators $A^\beta$ such that
\[\langle\beta_i|A= \langle I|_{\bar{i}i}A^{\beta_i}\otimes \openone_2.\]
ii/ Qubit operators can travel through singlets as
\[A^\beta\otimes \openone|I\rangle=\openone\otimes\sigma_y(A^{\beta})^T\sigma_y|I\rangle.\]
iii/ The concepts of quantum teleportation and entanglement
swapping \cite{tele,swapping} allow two unentangled particles,
each of them entangled to a different auxiliary particle, to
become entangled by doing a joint measurement on these auxiliary
particles in the Bell basis:
\[\left._{k\bar{k}}\langle I|\right.\left(|I\rangle_{\bar{i}k}\otimes|I\rangle_{\bar{k}j}\right)=-|I\rangle_{\bar{i}j}\]

The crucial observation is now that a von-Neumann measurement on
the $k$'th spin--1 in the basis
\be\label{basis}\{|\alpha_k\rangle\}=\{|0\rangle_k,|\pm\rangle_k\}\equiv\{|0\rangle_k,(|-1\rangle_k\pm|1\rangle_k)/\sqrt{2}\}\ee
exactly corresponds to a Bell measurement on the two qubits
$\bar{k}$ and $k$. If all spin 1's are therefore measured in this
local basis, the mechanism of entanglement swapping  will produce
a Bell state between the two qubits at the end points of the
chain, independent of its length. This proves the existence of an
infinite entanglement length in ground states of gapped spin
Hamiltonians.

To be more precise, the three properties above allow to rewrite
$|V\rangle$ in the very convenient matrix product form \cite{FNW},
from which all correlation functions and the LE can be calculated
exactly. Indeed, inserting a resolution of the identity
$\openone_{3^N}=\otimes_{i=1}^N(\sum_{\alpha_i}
|\alpha_i\rangle\langle\alpha_i|)$ in expression (\ref{VBS0}), it
follows that
\be\label{VBS}|V\rangle=\sum_{\alpha_1,\alpha_2,\ldots,\alpha_N}|\alpha_1\rangle\ldots
|\alpha_N\rangle\left(\openone\otimes A^{\alpha_N}\cdots
A^{\alpha_1}\right)|I\rangle_{\bar{0},N+1} \ee If one chooses the
basis (\ref{basis}), then $A^0=\sigma_z,A^+=\sigma_y,A^-=\sigma_x$
with $\sigma_\alpha$ the Pauli matrices.  The information about
the remaining Bell state after measuring the spins $1\ldots N$ can
of course be deduced from the measurement outcomes.

In the case of the AKLT-model, the so--called string order
parameter \cite{string} has been studied extensively because it
reveals a hidden topological long-range order. It is defined as
the expectation value of the multi-site observable
\be\label{stri}\langle \sigma^z_{\bar{0}}(\otimes_{k=1}^N
\exp(i\pi S^z_k))\sigma^z_{N+1}\rangle.\ee The existence of such
an order can easily be understood from the formalism presented
here. The operators $\exp(i\pi S^z_k)$ are all diagonal in the
basis (\ref{basis}). If one measures all spin 1's in that basis,
the fact that the final Bell state will be
$|00\rangle\pm|11\rangle$ or $|01\rangle\pm|10\rangle$ is solely
determined by the parity of the number of times $N_0$ the
measurement outcome is $0$; indeed, $A^{\alpha_N}\cdots
A^{\alpha_1}$ will be diagonal if and only if $N_++N_-=N-N_0$ is
even. The operator $\otimes\exp(i\pi S^z)$ is exactly keeping
track of this parity. Taking the trace can be interpreted as
averaging states after measurement, where $\otimes\exp(i\pi S^z)$
introduces a negative weight to the states with odd $N-N_0$.
Therefore the 2-qubit operator ${\rm Tr}_{1\ldots
N}[(\otimes_{k=1}^N \exp(i\pi S^z_k))|V\rangle\langle V|]$ has
diagonal elements $[1/4,-1/4,-1/4,1/4]$ and hence maximal
correlations in the zz-direction, independent of $N$. The string
order parameter is therefore a manifestation of the symmetries in
the mechanism of entanglement swapping.

\begin{figure}
\includegraphics[width=8cm]{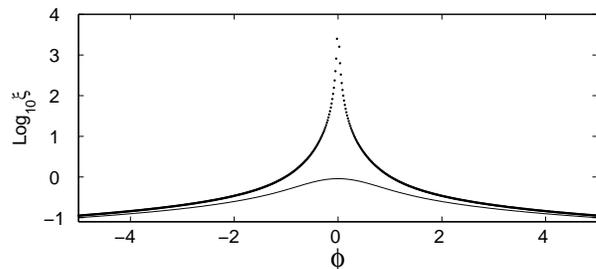}
\vspace{-15 pt} \caption{Correlation length $\xi_C$ (solid) and of
the Entanglement Length $\xi_E$ (dotted) for the ground state of
the deformed AKLT-Hamiltonians $H(\phi)$ (\ref{Hphi}).}
\label{figAKLT}
\end{figure}

Let us next investigate what will happen to the entanglement
length when the AKLT-Hamiltonian is deformed. We introduce the
following 1-parameter family of Hamiltonians:
\begin{eqnarray}
H(\phi)&=&\sum_{k=0}^N X_{k,k+1}(\phi)\label{Hphi}\\
X_{k,k+1}(\phi)&=&((\Sigma^\phi_k)^{-1}\otimes
\Sigma_{k+1}^\phi)X_{k,k+1}^{\rm AKLT}((\Sigma_k^\phi)^{-1}\otimes
\Sigma_{k+1}^\phi)\nonumber\end{eqnarray} where $\Sigma_k^\phi$ is
defined as
\[
\Sigma_{k}^\phi=\openone_k+(\cosh(\phi)-1)S^z_k+\sinh(\phi)(S^z_k)^2.\]
The AKLT-model corresponds to $\phi=0$, and the perturbation
breaks the $O(3)$ rotational symmetry to $O(2)$.
In the limit of $\phi\rightarrow\pm\infty$, the unique ground
state is a product state with all individual spins eigenstates of
$S^z_k$ with eigenvalue $0$. The unique ground state is completely
specified by replacing $A$ in (\ref{VBS0}) by
\[A=\left(%
\begin{array}{cccc}
  \exp(\phi) & 0& 0 & 0 \\
  0& \frac{\exp(-\phi)}{\sqrt{2}} & \frac{\exp(\phi)}{\sqrt{2}} & 0\\
  0 & 0 & 0& \exp(-\phi) \\
\end{array}%
\right).\] As will be explained later, one can calculate the
correlation functions and the LE exactly for the end points of the
chain. The associated correlation length $\xi_C$ and entanglement
length $\xi_E$ are given by
\begin{eqnarray*}
\xi_C&=&1/\ln\left|\sqrt{\cosh(2\phi)^2+3}+\cosh(2\phi)\right|\\
\xi_E&=&1/\ln\left|\frac{\sqrt{\cosh(2\phi)^2+3}+\cosh(2\phi)}{3}\right|\end{eqnarray*}
and shown in Figure \ref{figAKLT}. The quantity $\xi_C$ remains
always finite and attains its maximum for $\phi=0$. $\xi_E$ is
always strictly larger than $\xi_C$, indicating the presence of
two different length scales in the ground state. Moreover, $\xi_E$
diverges at the AKLT-point $\phi\rightarrow 0$. This quantum
transition with a diverging length scale remains clearly
undetected by the properties of the correlation functions.

The optimal measurement basis turns out to be independent of
$\phi$ and given by (\ref{basis}). The reason that $\xi_E$ is
finite for $\phi\neq 0$ is the fact that the perturbation
effectively replaces the singlet $|I\rangle=|01\rangle-|10\rangle$
in equation (\ref{VBS0}) with a non-maximally entangled state
$\exp(\phi)|01\rangle-\exp(-\phi)|10\rangle$ (see
Fig.~\ref{fig0}); using such a state for entanglement swapping
degrades the entanglement (a \emph{bias} is added at each step),
hence giving rise to an exponential decay and a finite
entanglement length.

Let us now develop the mathematical formalism to calculate
correlation functions and the LE. We will consider a
generalization of the AKLT-states, the family of so--called
finitely correlated states (FCS) \cite{FNW}, which are, in the
appropriate limit, dense in the subspace of all translational
invariant states. These FCS are completely parameterized by a
matrix $A$ as in equations (\ref{VBS0},\ref{VBS}), but instead of
spin--1 systems we consider general spin--$S$ systems. The
spin--$1/2$'s are replaced by a spin--$(D-1)/2$, and $|I\rangle$
becomes a maximally entangled state in a $D\times D$ Hilbert space
[$A$ is a $(2S+1)\times D^2$ matrix; note that $S$ and $D$ are two
independent parameters]. These are all unique ground states of
gapped local Hamiltonians that can be constructed by calculating
the orthogonal complement of an expression equivalent to
(\ref{proj}). For simplicity, we will again consider a spin chain
with spin $S$ at sites $1\ldots N$ and spin $(D-1)/2$ at the end
points.
Expectation values of the form $\langle V|\hat{O}_1\otimes
\hat{O}_2\otimes \cdots \hat{O}_N|V\rangle$ can readily be
calculated by defining the real $D^2\times D^2$ matrices
$R(\hat{O})$ as \be\label{R}R_{ij}(\hat{O})=M_{ik}{\rm
Tr}\left[(A^\dagger \hat{O}A)\sigma_j\otimes\sigma_k\right]\ee
with $\{\sigma_\alpha\}$ a complete orthonormal basis for
hermitian operators including $\sigma_0=\openone$. The matrix $M$
depends on the choice of $|I\rangle$ and of the basis
$\{\sigma_\alpha\}$; in the case of the singlet state and
$\{\sigma_\alpha\}$ the Pauli matrices, $M={\rm
diag}[1,-1,-1,-1]$. For the example presented in the previous
section, e.g., $R(\hat{\openone})$ is given by
\[
R(\openone)=\left(%
\begin{array}{cccc}
  3\cosh(2\phi) & 0 & 0  & \sinh(2\phi) \\
  0 & -1 & 0 & 0 \\
  0 & 0 & -1 & 0 \\
  -3\sinh(2\phi) & 0 & 0 & -\cosh(2\phi) \\
\end{array}%
\right).\] Evaluating expectation values becomes equivalent to
calculating the $(0,0)$ element of the matrix product
\[\left[R(\hat{O}_N)R(\hat{O}_{N-1})\cdots R(\hat{O}_1)\right]_{(0,0)}.\]
The normalization of $|V\rangle$ is clearly given by
$\left[(R(\openone))^N\right]_{(0,0)}$, and therefore correlation
functions can be calculated as
\[\langle\hat{O}_1\otimes \hat{O}_2\otimes \cdots
\hat{O}_N\rangle=\frac{\left[R(\hat{O}_N)\cdots
R(\hat{O}_1)\right]_{(0,0)}}{\left[(R(\openone))^N\right]_{(0,0)}}.\]
To calculate the LE, we need to do an optimization over all
possible local bases $\{\beta_i\}$ (including overcomplete ones
corresponding to generalized measurements). If one measures all
spins $k=1\ldots N$ in the basis $\{|\beta_k\rangle\}$, the state
of the two extremal spins is conditioned on the outcomes
$\{\beta_1,\ldots,\beta_N\}$ and given by \be\label{2q}
|\psi_{\beta_1,\ldots,\beta_N}\rangle=\left(\openone\otimes
A^{\beta_N}\cdots A^{\beta_1}\right)|I_{\bar{0},N+1}\rangle, \ee
where we used the notation $\langle\beta_i|A=\langle
I|A^{\beta_i}\otimes\openone$. The probability for this outcome to
happen is
\[p_{\beta_1,\ldots,\beta_N}=\frac{\langle\psi_{\beta_1,\ldots,\beta_N}|\psi_{\beta_1,\ldots,\beta_N}\rangle}{\langle
V|V\rangle}.\] Following \cite{VDD03}, a generalization of the
concurrence \cite{Wootters} for pure bipartite $D\times D$ states
is given by $C(|\chi\rangle=B\otimes
\openone_D|I\rangle)=|det(B)|^{2/D}/\langle\chi|\chi\rangle$.
Using this measure, the average entanglement factorizes and is
given by
\[\frac{\sum_{\beta_1,\ldots,\beta_N}|\det(A^{\beta_N}\cdots
A^{\beta_1})|^{\frac{2}{D}}}{\langle
V|V\rangle}=\frac{\prod_i\left(\sum_{\beta_i}
|\det(A^{\beta_i})|^{\frac{2}{D}}\right)}{\langle V|V\rangle}\]
Obviously, the optimal strategy is to measure all spins in the
same basis that maximizes $\sum_{\beta_i}
|\det(A^{\beta_i})|^{\frac{2}{D}}$. This problem is equivalent to
calculating the entanglement of assistance (EoA) \cite{EoA} of the
(unnormalized) $D^2\times D^2$ state $A^\dagger A$, and can in
general easily be done numerically.

In the case of singlet valence bonds ($D=2$) and $A$ an arbitrary
$3\times 4$ matrix, this EoA can be calculated exactly
\cite{laustsen}. Making use of the results in \cite{lorsvd}, one
obtains the exact expression for the LE: \be\label{LE}
E_{\bar{0},N+1}=\frac{\left[\sqrt{\lambda_{\max}(MR(\openone)M(R(\openone))^T
)}\right]^{N}}{\left[(R(\openone))^N\right]_{00}}.\ee Here
$R(\openone)$ was defined in (\ref{R}), $M={\rm
diag}[1,-1,-1,-1]$, and $\lambda_{\max}$ means the largest
eigenvalue.

Let us investigate the general conditions under which the
entanglement length as defined by the LE (\ref{LE}) diverges. It
can be shown \cite{Popp} that this will happen if and only if
there exists an operator $Q$ such that the EoA of $(Q^\dagger
\otimes Q^\dagger )A^\dagger A(Q\otimes Q)$ is maximal, i.e. when
it can be written as a convex combination of maximally entangled
states (such as for the AKLT); this is indeed the necessary and
sufficient condition for perfect entanglement swapping to be
possible. Let us compare this with the condition for the presence
of a generalized string order parameter, which we define as in
(\ref{stri}) but with an extra optimization over the hermitian
operator $X$ in $\otimes_k \exp(i\pi X_k)$ instead of $S_z$. The
condition is now the existence of an operator $Q$ such that
$(Q^\dagger \otimes Q^\dagger )A^\dagger A(Q\otimes Q)$ is
diagonal in a basis of maximally entangled states that have either
support in the $|01\rangle\pm|10\rangle$ or in the
$|00\rangle\pm|11\rangle$ subspace; this gives two extra
nontrivial constraints as compared to the condition for a
diverging entanglement length, and hence there exist ground states
with vanishing string order parameter, but infinite entanglement
length.

Let us conclude by discussing some generalizations to the present
results. First of all, the formalism developed in this paper
allows us to calculate the localizable entanglement for ground
states of arbitrary Hamiltonians numerically by making use of the
Density Renormalization Group formalism (DMRG) \cite{DMRG}: the
fixed point of the DMRG algorithm yields a FCS \cite{Ostlund,FNW}
on which we can apply the machinery presented. Secondly, the
results presented equally apply to higher dimensional AKLT-models,
and reveal an intriguing connection with quantum computation
\cite{cluster}. Finally, we have done numerical diagonalizations
of the Heisenberg antiferromagnetic spin-1 Hamiltonian
$H=\sum_k\vec{S}_k\vec{S}_{k+1}$ with spin 1/2's at the end points
\cite{Popp}. The localizable entanglement between the end points
is again $1$ independent of the number of spins $N$ and is
obtained by measuring in the optimal basis (\ref{basis}) for the
AKLT: the entanglement length in the Heisenberg spin-1
antiferromagnet is also infinite, and hence this ground state
could be used as a perfect quantum channel. It is interesting to
note that the entanglement length for the spin 1/2 antiferromagnet
is also infinite \cite{Korepin}: the presence of a Haldane-gap
severely affects the correlation length, but does not seem to
affect the entanglement length.

This work has been supported by the EU IST program (QUPRODIS and
RESQ), \emph{Kompetenznetzwerk Quanteninformationsverarbeitung der
Bayerischen Staatsregierung}, SFB631, and DGES under contract
BFM2000-1320-C02-01. We thank M. Fannes for useful discussions.

\end{document}